\documentstyle[emulateapj,apjfonts,epsfig,rotating,graphicx]{article}

\slugcomment{}

\begin{document}

\title{On the cross correlation between the arrival direction of ultra-high
energy\\ cosmic rays, BL Lacertae, and EGRET detections: A new way to
identify EGRET sources?}

\author{Diego F. Torres$^a$,
Stephen Reucroft$^b$, Olaf Reimer$^c$, and Luis A.
Anchordoqui$^b$} \affil{$^a$ Lawrence Livermore National
Laboratory, 7000 East Ave. L$-$413, Livermore, CA 94550, USA\\
dtorres@igpp.ucllnl.org } \affil{$^b$ Department
of Physics, Northeastern University, Boston, MA 02115, USA\\
stephen.reucroft@cern.ch, l.anchorodoqui@neu.edu } \affil{$^c$
Institut f\"ur Theoretische Physik, Ruhr-Universit\"at Bochum,
44780 Bochum, Germany\\ olr@tp4.ruhr-uni-bochum.de}

\begin{abstract}
With the aim of testing recent claims for a particularly strong
correlation between ultra-high energy cosmic rays (UHECRs),
observed with the AGASA  and the Yakutsk
experiments, and a sample of  BL Lacertae (BL  Lacs),
we here conduct a blind statistical assessment. We
search for associations between the same set of BL Lac objects
and the arrival directions of 33 relevant UHECRs observed with the
Haverah Park and the Volcano Ranch experiments. Within the accuracy of
angle determination, there are no positional coincidences. The
probability that this null result arises as a statistical
fluctuation from the strongly correlated case is less than 5\%.
This implies that the possible correlation between the arrival
directions of UHECRs and BL Lacs is not
statistically sustained.  We discuss the impact of our findings on
the propose additional connection among UHECRs, BL Lacs, and EGRET
$\gamma$-ray blazars. Recently, such an association
was used as  classification technique for EGRET sources. Here we
show that its main underlying hypothesis, i.e., the EGRET angular uncertainty
is twice that quoted in the Third EGRET Catalog, grossly underestimates the goodness
of existing gamma ray data.

\end{abstract}

\keywords{Ultra-high energy cosmic rays  -- BL Lacertae -- gamma rays \hspace{2cm} NUB-3238-TH-03 }

\section{Introduction}

Farrar and Biermann (1998) pointed out the existence of a
directional correlation between compact radio quasars (QSOs) and
ultra-high energy cosmic rays (UHECRs). Their claim was supported
by two significant factors: (i) there was an {\it a priori}
postulated theoretical reason by which to expect such an
alignment, i.e. the existence of a new neutral hadron that would
travel unscathed all the way to the Earth (Farrar 1996, Chung et
al. 1998) or neutrinos producing $Z$-bursts (Weiler 1999; Fargion,
Mele \& Salis 1999; Fodor, Katz \& Ringwald 2002), and (ii) all
events at the high end of the spectrum observed by that time, with
energy at least 1$\sigma$ above $10^{19.9}$~eV, were aligned with
high redshifted quasars, a phenomenon with a chance probability of
occurrence less than 0.5\% (Farrar \& Biermann 1998). This report
quickly opened a large debate on whether UHECRs can evade
interactions with cosmic microwave background photons, and arrive
at Earth evading the Greisen (1966) -- Zatsepin--Kuzmin (1966),
GZK, cutoff. Most of the alternative explanations to evade the GZK
cutoff require physics beyond the standard model (for an exception
see Kalashev et al. 2001), including violation of local Lorentz
invariance (Coleman \& Glashow 1998), photon-axion mixing (Csaki
et al. 2003), and neutrinos interacting strongly in the atmosphere
(due to physics beyond the --perturbative-- Standard Model, Fodor
et al. 2003; or because of TeV-scale gravity, Domokos \&
Kovesi-Domokos 1999, Jain et al. 2000). The latter, however, is
severely constrained by observations (Anchordoqui et al. 2001). In
this {\it Letter}, we first comment on the status of the
correlation between QSOs and UHECRs, and then analyze more recent
strong claims for a correlation between UHECRs and BL Lacertae (BL
Lacs), a subgroup of the previously studied QSOs. Finally, we
scrutinize a newly proposed classification technique of EGRET
sources, based on the cross correlation of BL Lacs, UHECRs, and
$\gamma$-ray catalogs.

\section{QSOs and UHECRs}

The possible correlation between UHECRs and QSOs was subject to a
great deal of scrutiny. Hoffman (1999) stated that one of the 5
events used in the Farrar and Biermann (1998) study, the highest
energy event observed by the Fly's Eye experiment (Bird et al.
1995), should not be included in the UHECR sample under analysis,
because this very same event was previously considered to
introduce the hypothesis. Without this event, the positive
alignment with random background probability is increased to $
<3\%$ (Farrar \& Biermann 1999). Using an updated event list
(twice the size of the previous) from the Haverah Park (Ave et al.
2000) and the AGASA (Hayashida et al. 2000) experiments, Sigl et
al. (2001) showed that the statistical significance of the
alignment is lowered to 27\%. More recently, Virmani et al. (2002)
favored the earlier proposed alignment. However, it should be
stressed that most of the Virmani et al. correlation signal comes
from events with large uncertainty both in energy and in position:
they considered events from the SUGAR experiment, but it is not
clear whether these events were above the GZK cut-off (see, e.g.,
Anchordoqui et al. 2003).

Very recently, the Haverah Park energy estimates have been
re-assessed (Ave et al. 2003). For the cosmic rays in question,
the energy of the 2 events observed by this array with incident
zenith angle $<45^\circ$, that was previously quoted as $>
10^{19.9}$~eV at 1$\sigma$, is now shifted $\approx 30\%$
downwards, below the energy cut chosen by Farrar and Biermann
(1998). Hence, independently of the statistical test used, when
considering only the highest energy ($> 10^{19.9}$ eV at
1$\sigma$) events\footnote{Those events would be most interesting
for new physics, because they have no contamination from the
expected proton pile-up around the photopion production
threshold.} the correlation between UHECRs and QSOs is consistent
with a random distribution at the $1\sigma$ level.

\section{BL LACs and UHECRs}

In a series of recent papers, Tinyakov and Tkachev (2001, 2002,
2003) claim a correlation between the arrival directions of UHECRs
and BL Lacs, a subgroup of the QSO sample previously considered.
Specifically, the BL Lacs chosen were those identified in the
(9th-Edition) Veron-Cetty and Veron (2000) catalogue of Quasars
and Active Galactic Nuclei, with redshift $z > 0.1$ or unknown,
magnitude $m < 18$, and radio flux at 6 GHz $F_6 >
0.17$~Jy.\footnote{The catalogue of Quasars and Active Galactic
Nuclei is regularly updated, see Veron-Cetty and Veron (2001) for
the 10$^{\rm th}$-Edition. The 9th-Edition is electronically
available at {\tt
http://www.obs-hp.fr/www/catalogues/veron2$\_$9/veron2$\_9$.html.}}
Only 22 objects fulfill such restrictions. In this analysis there
is no buffer against contamination by mismeasured protons piled up
at the GZK energy limit. The CR sample of Tinyakov and Tkachev
consists of 26 events measured by the Yakutsk experiment with
energy $> 10^{19.38}$~eV  (Afanasiev et al. 1996), and 39 events
measured by the AGASA experiment with energy $> 10^{19.68}$~eV
(Hayashida et al. 2000). The evidence supporting their claim is
based on 6 events reported by the AGASA Collaboration (all with
average energy $< 10^{19.9}$~eV), and 2 events recorded with the
Yakutsk experiment (both with average energy $< 10^{19.6}$~eV),
which were found to be within $2.5^\circ$ of 5 BL Lacs contained
in the restricted sample of 22 sources. The chance probability for
this coincidence set-up is found to be $2 \times 10^{-5}.$

One drawback of the claim made by Tinyakov and Tkachev (2001) is
that the data set used to make the initial assertion is also being
used in the hypothesis testing phase. Note that if enough searches
are performed on a finite data set which is sampled from an
isotropic distribution, some highly significant positive results
are certain to occur due to the statistical fluctuations that
necessarily arise in any finite sampling. Evans, Ferrer and Sarkar
(2002) already called into question whether the selection criteria
for the subset of brightest BL Lacs are unbiased. Strictly
speaking, Tinyakov and Tkachev imposed arbitrary cuts on the BL
Lac catalogue so as to maximize the signal-to-noise ratio,
compensating {\it a posteriori} the different cut adjustments by
inclusion of a penalty factor. Without these arbitrary cuts, the
significance of the correlation signal is reduced at the 1$\sigma$
level (Evans, Ferrer \& Sarkar 2002). Moreover, even in acceptance
of this {\it a posteriori} approach, the estimated value of the
penalty factor is subject to debate 
(Evans, Ferrer \& Sarkar 2002; Tinyakov \& Tkachev 2003).

Given the pivotal role played by the penalty factor in testing the
hypothesis with a single set of data, it is of interest to
circumvent this ambiguity by
performing a blind analysis. We have at our disposal the
cosmic ray arrival directions of the Haverah Park (Stanev et al.
1995) and Volcano Ranch (Linsley 1980) experiments, which,
although not useful to distinguish a positive correlation (because
the penalties involved are probably already as large as the signal
which one expects to test), they provide the framework to
disregard the correlation if none is found in the data.

Surface arrays in stable operation have nearly continuous
observation over the entire year, yielding a uniform exposure in
right ascension. However, the declination distribution is
different for each experiment, because the relative efficiency
of the detection of events depends upon the latitude of the array
and detector type. As shown by Uchihori et al. (2000), the field
of view of AGASA + Yakutsk is roughly equal to that of Volcano
Ranch + Haverah Park. It is noteworthy that even though the energy
of the Haverah Park events has been reduced by about 30\% (Ave et
al. 2003), the 27 events contained in our sample, originally with
energy $> 10^{19.6}$~eV  (Lawrence, Reid \& Watson 1991), are well
above the energy cut for Yakutsk's events selected by Tinyakov
and Tkachev. Combined with the 6 events recorded at the Volcano
Ranch with energy $> 10^{19.6}$~eV  (Linsley 1980), we have a
virgin data-set of 33 events, amounting to half of the cosmic-ray
arrival directions used to make the claim.

\begin{figure*}
\centering
\includegraphics[width=.75\textwidth,height=7.5cm]{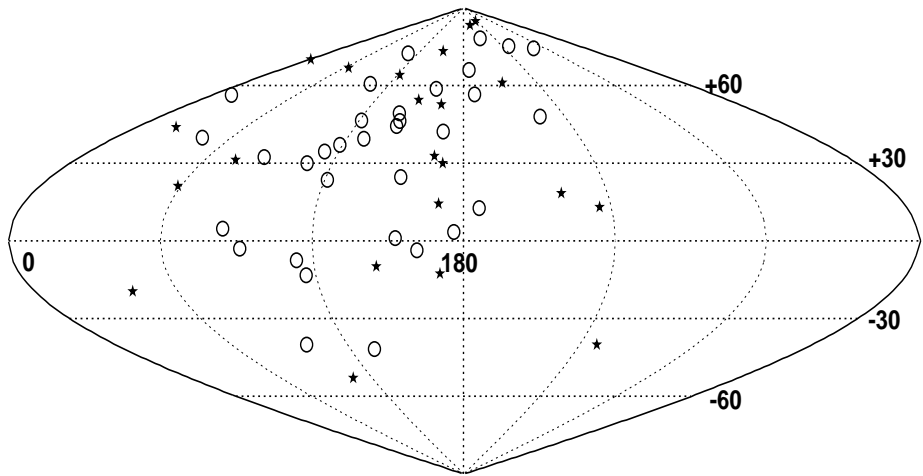}
\caption{The open circles indicate the arrival direction (in
galactic coordinates, $l$ and $b$) of 33 UHECRs with incident
zenith angle $<45^\circ$ observed by the Haverah Park (27 events)
and the Volcano Ranch (6 events) arrays. There are two sets of CRs
clustering within experimental angular resolution (Uchihori et al.
2000). Namely, a Haverah Park doublet with coordinates $(l=
140.98^\circ,\, b=49.43^\circ) \,\, +\,\,
(l=143.60^\circ,\,b=46.30^\circ)$ and a mix-doublet Volcano Ranch
$(l = 143.00^\circ,\, b=44.30^\circ)$ + Haverah Park
$(l=143.60^\circ,\,b=46.30^\circ)$. The stars stand for the 22 BL
Lacs from the 9th-Edition of the Veron-Cetty and Veron (2000)
catalogue of Quasars and Active Galactic Nuclei, with redshift $z
> 0.1$ or unknown, magnitude $m < 18$, and radio flux at 6 GHz
$F_6 > 0.17$~Jy.} \label{skymap}
\end{figure*}

In Fig.~\ref{skymap} we plot the position on the sky in galactic
coordinates of both the UHECRs and the selected BL Lacs. There
are no positional coincidences between these two samples up to an
angular bin $> 5^\circ$. Such an angular scale is well beyond the
error in arrival determination, which is found to be $\approx
3^\circ$ (Uchihori et al. 2000). On the basis of the strongly
correlated sample analyzed by Tinyakov and Tkachev, one expects
the distribution describing the correlation between the set of BL
Lacs and any UHECR data-set with 33 entries to be Poisson with
mean $\approx 4.06.$
Taking the data at face value, this implies a
$2\sigma$ deviation effect. Moreover, the 95\% CL interval of the
distribution which samples the correlation between the BL Lacs and
cosmic rays recorded by Volcano Ranch + Haverah Park is (0, 3.09)
(see, e.g. Feldman \& Cousins  1998). Therefore, the probability
to measure the expected mean value $\approx 4.06$ is $\ll 5\%$.
All in all, the 8 coincidences in the Tinyakov and Tkachev (2001)
analysis do not represent a statistically significant effect.

\section{UHECRs and EGRET AGNs}

On a similar track, Gorbunov et al. (2002) claimed that a set of
$\gamma$-ray loud BL Lac objects can be selected by intersecting
the EGRET and BL Lacs catalogs. The only requirement that Gorbunov
et al. considered for a BL Lac to be physically associated with an
EGRET source is that the angular distance between the best
estimated position of the pair does not exceed $2 R_{95}$, where
$R_{95}$ is the 95\% CL contour of the EGRET detection.

Their claim was based on a positional correlation analysis (using
the doubled size for EGRET sources) between the Third EGRET
Catalog (3EG, Hartman et al. 1999) and the objects identified as
BL Lac in the Veron-Cetty \& Veron (2000) Catalog. This results in
14 coincidences, 4 of which are further found to be part of the 5
BL Lacs located within 2.5$^\circ$ of UHECRs discussed above.

The typical $R_{95}$ radius for EGRET sources is 0.5--1$^\circ$.
Because of such large uncertainties, a standard practice in
$\gamma$-ray studies aiming to give preliminary associations
between EGRET sources and possible counterparts is to study, in
addition to the object being proposed, any other coincident system
able to generate photons in the EGRET range (100~MeV--10~GeV). All
of the latter should be discarded as the origin of the high energy
radiation in order for the association claim to persist. This
process usually involves theoretical modelling and multiwavelength
observations (see e.g. Caraveo 2002, Reimer et al. 2001, Torres et
al. 2003a, and references therein).

\begin{figure*}[t]
\centering
\includegraphics[width=.32\textwidth,height=5.5cm]{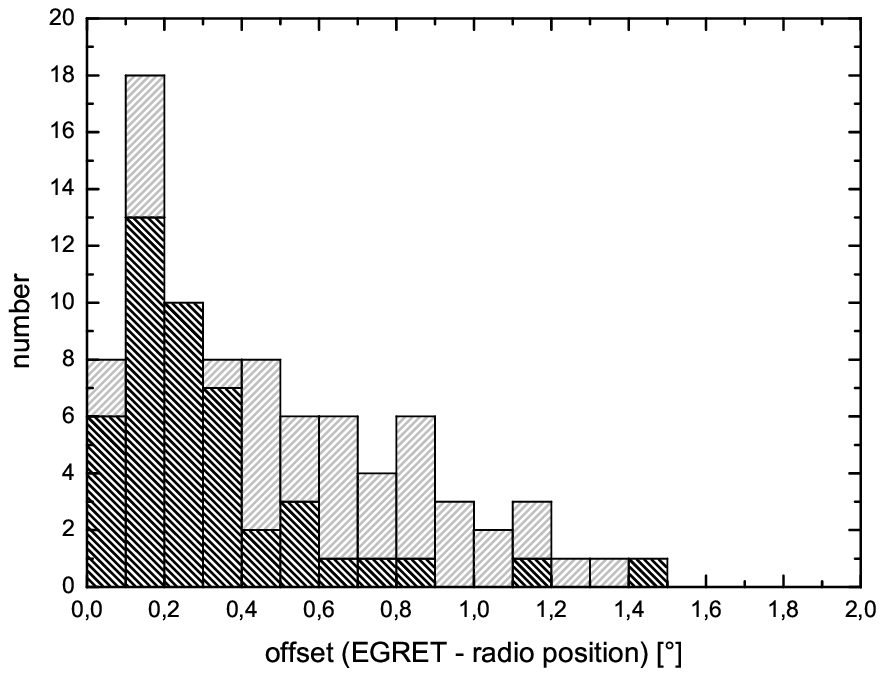}
\includegraphics[width=.32\textwidth,height=5.5cm]{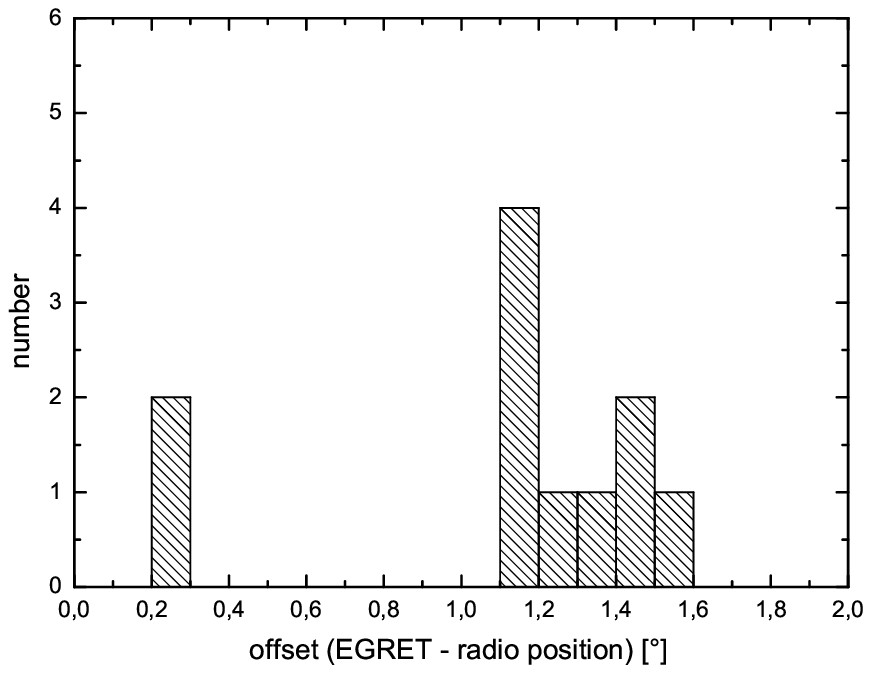}
\includegraphics[width=.32\textwidth,height=5.5cm]{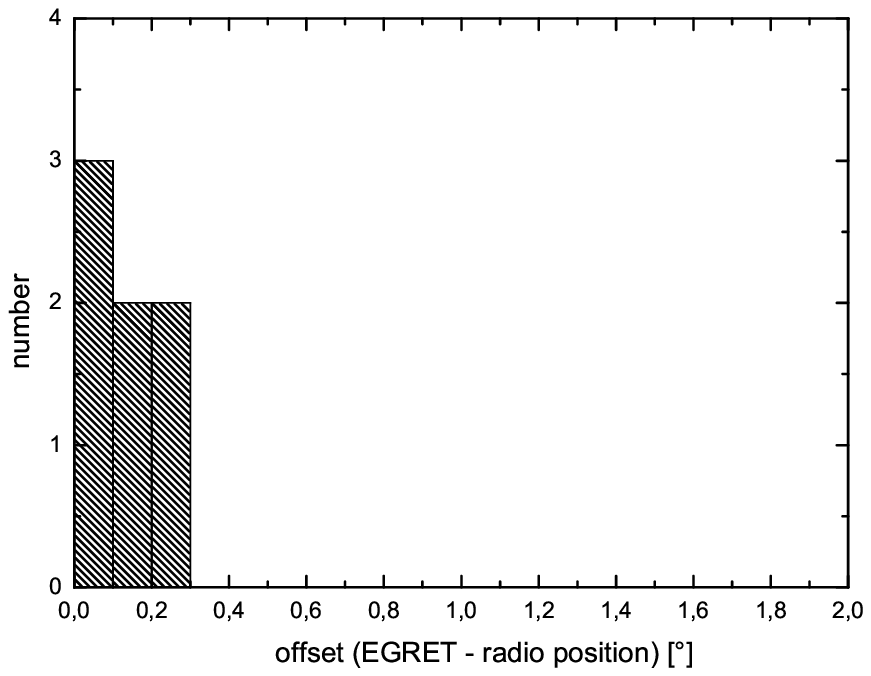}
\caption{Left: Offset distribution between $\gamma$-ray and radio
position of high-confidence [dark histogram] and plausible [light histogram]
 3EG-classified AGNs. Middle: Offset distribution for the newly proposed associations using the
 Gorbunov at al. technique (see text). Right: Offset distribution for 3EG
 pulsars.
 } \label{olaf_f1}
\end{figure*}

The case of active galactic nuclei (AGNs) as EGRET counterparts
has been analyzed by Mattox, Hartman and Reimer (2001), who
provided a spatial-statistical
assessment.
They list 46 EGRET sources with high probability of
being identified with known blazars, confirming 45 AGN identifications
made in the 3EG (Hartman et al. 1999). A further 39
EGRET sources have been listed as plausible AGN identifcations.
In Fig.~\ref{olaf_f1} (left panel) we show the positional offset between
the maximum-likelihood-algorithm for the $\gamma$-ray
source position  (Mattox et al. 1996) and the radio position of the
respective AGN
identification. We show separately those AGN considered as high-confidence and plausible
counterparts.

Independent support for some of these AGN identifications results from
dedicated
multifrequency counterpart observations, including spectroscopic
confirmation of
their blazar nature (Sowards-Emmerd, Romani \& Michelson 2003,
Halpern, Eracleous \& Mattox 2003).
In contrast to the offset distribution between $\gamma$-ray and radio
position
of confirmed (or at least, most probable)
AGN identifications, the offset distribution of the newly
suggested
AGN counterparts by Gorbunov et al. (2002), shown in Fig.~\ref{olaf_f1} (middle panel),
has a completely different shape. We attribute this to the inappropriate consideration of source
localization
uncertainties of EGRET-detected $\gamma$-ray sources, which lead Gorbunov
et al. to suggest counterparts well into the range of $2 R_{95}$ of an individual
EGRET source. This extension of the EGRET angular uncertainty was motivated in
a discrepancy between the radio and the
$\gamma$-ray position of the Vela pulsar, which will be discussed below. For
comparison, Fig.~\ref{olaf_f1} (right panel) gives the offset distribution for the radio and
$\gamma$-ray positions
of identified pulsars in the EGRET data. Clearly, the range of absolute
offset values,
based on phase selected $\gamma$-ray events, is minimal in the case of the
pulsars.

The 3EG source related to Vela is a very special
case. It is the strongest known $\gamma$-ray source, and one of the best
localized, $R_{95} = 0.021^\circ$. The 95\% CL contour of the EGRET
detection and the offset with Vela are
both one order of magnitude less than the typical values of these
quantities in the 3EG. The misplacing for Vela occurs because  the
analysis technique privileges the discovery and correct detection
of weaker sources, and it is applied to all EGRET sources in the
3EG (2/3 of which are unidentified with no obvious candidates) identically.
The offset of the Vela position and, in general, of bright
sources, is minimized by using map bins smaller than the standard
0.5$^\circ$ used in the 3EG. This increases the computation time
greatly; and since all of the most significant sources were
identified with objects whose positions were well known, the
smaller bin size was not adopted to give source positions in the
3EG.  See the comment on the source 3EG J0834-4511 (Vela) in the
section of particular detections of the 3EG Catalog (Hartman et
al. 1999). Systematics, then, do not pose a major problem for the
source location capability of EGRET, even in regions of
significant diffuse emission or strong nearby sources (Hartman et
al. 1999). Most importantly, the error contours for many of the
AGNs show that the location capability improves
for regions away from the Galactic plane, where most of the
blazars are.

In addition, some of the 3EG-associated AGNs could be false
positives (i.e. AGNs that are mis-associated with EGRET sources by
a failure of the statistical methods used in the classification).
This fact is particularly important for statistical methods based
{\it only} on the relative positions between the candidate and the
EGRET source center (see Torres 2003b for a review). Working with
114 sources above $|b|>10^{{\rm o}}$, Punsly (1997) has estimated
the number of random coincidences as a function of the field
radius: $\sim 2$ (10) quasars with more than 1 Jy of 5 GHz flux
are expected to correlate by random chance if the size of the
typical EGRET angular uncertainty is 0.7$^{{\rm o}}$ (1.7$^{{\rm
o}}$). The number of random coincidences increases as the
radio-loudness of the AGN decreases (since there are more AGNs
with smaller flux). This sheds additional doubt on the
correlations found beyond the 95\% location contours of EGRET
sources.

\section{Concluding Remarks}

Available statistics on the arrival directions of the UHECRs
reveals no significant correlations above random with BL Lacs nor
with any other type of quasars, including EGRET blazar detections.
Furthermore, identifying EGRET sources with BL Lacs just by
positional pairing within twice the EGRET error grossly
underestimates the goodness of existing gamma-ray data.
\footnote{In closing, we comment on the apparent correlation
between nearby dormant quasar remnants (``dead quasars'') and the
arrival directions of the 38 UHECRs measured by the AGASA
experiment with energy $> 10^{19.6}$~eV and $|b| > 20^\circ$
(Torres et al. 2002). 
We note that NGC 2300 (which is
the only quasar remnant hosted in a non-elliptical galaxy) is
actually beyond the reach of the AGASA experiment, which improves
a little the correlation presented by Torres et al. (2002). 
Since there are 25 events with $|b| > 20^\circ$ (some of them with
energy $< 10^{19.6}$~eV) in the data sample which combines
measurements of the Volcano Ranch and the Haverah Park
experiments, the dead quasar hypothesis (Boldt \& Ghosh 1999) is
limited by the statistics of small numbers (the expected mean
value being $\approx 1.31$ is just on the verge of the 68.27\% CL
interval, Feldman \& Cousins 1998), and the hypothesis awaits
further testing with larger sets of data.}

\acknowledgements

We gratefully acknowledge Elihu Boldt, Francesc Ferrer, Haim
Goldberg,  Tom McCauley, Andreas Ringwald, Subir Sarkar, Peter
Tinyakov, and Alan Watson, for useful discussions/email
correspondence. We further acknowledge Alan Watson for his kind
permission to use the Haverah Park data. The work of D.F.T. was
performed under the auspices of the U.S. D.O.E. (NNSA), by
University of California Lawrence Livermore National Laboratory
under contract No. W-7405-Eng-48. The research of S.R. and L.A.A
was partially supported by the US National Science Foundation
(NSF) under grant No. PHY-0140407.


\begin{thebibliography}{}

\bibitem{Afanasiev} Afanasiev B. N. et al., 1996 {\it in Proc. Int. Symp. on Extremely High Energy Cosmic
Rays : Astrophysics and Future Observatories} (ed. M.Nagano, Institute for Cosmic Ray Research,
Univ. of Tokyo) p.32.

\bibitem{Anchordoqui:2000uh}
Anchordoqui L., Goldberg H., McCauley T., Paul T., Reucroft S. \& Swain J., 2001
Phys.\ Rev.\ D {\bf 63}, 124009.


\bibitem{Anchordoqui:2002hs}
Anchordoqui L., Paul T., Reucroft S, \& Swain J., 2003
Int.\ J.\ Mod.\ Phys.\ A {\bf 18}, 2229.


\bibitem{Ave:2000nd}
Ave M., Hinton J. A., Vazquez R. A., Watson A. A., \& Zas E., 2000
Phys.\ Rev.\ Lett.\  {\bf 85}, 2244.

\bibitem{Ave:2001hq}
Ave M., Knapp J., Lloyd-Evans J., Marchesini M. \& Watson A. A., 2003
Astropart.\ Phys.\  {\bf 19}, 47.


\bibitem{Bird:1994uy}
Bird D. J. et al., 1995
Astrophys.\ J.\  {\bf 441}, 144.


\bibitem{Boldt:1999ge}
Boldt E. and Ghosh P., 1999
Mon.\ Not.\ Roy.\ Astron.\ Soc.\  {\bf 307}, 491.



\bibitem{Caraveo} Caraveo P. A. 2002, astro-ph/0206236,
To appear in the proceedings of the XXII Moriond Astrophysics
Meeting "The Gamma-Ray Universe" (Les Arcs, March 9-16, 2002),
eds. A. Goldwurm, D. Neumann, and J. Tran Thanh Van, The
GioiPublishers

\bibitem{Chung:1997rz}
Chung D. J., Farrar G. R., \& Kolb E. W., 1998
Phys.\ Rev.\ D {\bf 57}, 4606.


\bibitem{Coleman:1998en}
Coleman S. R. \& Glashow S. L., 1998
arXiv:hep-ph/9808446.


\bibitem{Csaki:2003ef}
Csaki C., Kaloper N., Peloso M. \& Terning J., 2003
arXiv:hep-ph/0302030.


\bibitem{Domokos:1998ry}
Domokos G. \& Kovesi-Domokos S., 1999
Phys.\ Rev.\ Lett.\  {\bf 82}, 1366.






\bibitem{Evans:2002jy}
Evans W., Ferrer F. \& Sarkar S., 2002
arXiv:astro-ph/0212533.


\bibitem{Fargion:1997ft}
Fargion D., Mele B. \& Salis A., 1999
Astrophys.\ J.\  {\bf 517}, 725.

\bibitem{Farrar:1996rg}
Farrar G. R., 1996
Phys.\ Rev.\ Lett.\  {\bf 76}, 4111.


\bibitem{Farrar:1998we}
Farrar G. R. \& Biermann P. L., 1998
Phys.\ Rev.\ Lett.\  {\bf 81}, 3579.


\bibitem{Farrar:1999fw}
Farrar G. R. \& Biermann P. L., 1999
Phys.\ Rev.\ Lett.\  {\bf 83}, 2472.




\bibitem{Feldman:1997qc}
Feldman G. J. \& Cousins R. D., 1998
Phys.\ Rev.\ D {\bf 57}, 3873.


\bibitem{Fodor:2001qy}
Fodor Z., Katz S. D. \& Ringwald A., 2002
Phys.\ Rev.\ Lett.\  {\bf 88}, 171101.


\bibitem{Fodor:2003bn}
Fodor Z., Katz S. D., Ringwald A. \& Tu H., 2003
Phys.\ Lett.\ B {\bf 561}, 191.





\bibitem{Gorbunov:2002hk}
Gorbunov D. S., Tinyakov P. G., Tkachev I. I. \&  Troitsky S. V., 2002
Astrophys.\ J.\  {\bf 577}, L93.





\bibitem{Greisen:1966jv}
Greisen K, 1966
Phys.\ Rev.\ Lett.\  {\bf 16} 748.


\bibitem{Halpern}Halpern J.P., Eracleous M.
\& Mattox J.R. 2003, Astron. J.  {\bf 125},  572.


\bibitem{Hartman} Hartman R. C. et al., 1999
Astrophys. J. Suppl. {\bf 123}, 79.



\bibitem{Hayashida:2000zr}
Hayashida  N. {\it et al.}, 2000
arXiv:astro-ph/0008102.



\bibitem{Hoffman:1999ev}
Hoffman C. M., 1999
Phys.\ Rev.\ Lett.\  {\bf 83}, 2471.

\bibitem{Jain:2000pu}
Jain P., McKay D. W., Panda S. \& Ralston J. P., 2000
Phys.\ Lett.\ B {\bf 484}, 267.

\bibitem{Kalashev:2001qp}
Kalashev O. E., Kuzmin V. A., Semikoz D. V. \& Tkachev I. I., 2001
arXiv:astro-ph/0107130.




\bibitem{Lawrence:cc}
Lawrence M. A., Reid R. J., \&  Watson A. A., 1991
J.\ Phys.\ G {\bf 17} 733.



\bibitem{Linsley} Linsley J., 1980 {\it Catalog of Highest Energy Cosmic Ray,} (World Data Center of
Cosmic Rays, Institute of Physical and Chemical Research, Itabashi, Tokyo) p.3.


\bibitem{M} Mattox, J. R. {\it et al.,} 1996
Astrophys. J. {\bf 461}, 396.



\bibitem{Mattox} Mattox J. R., Hartman R. C. \& Reimer O., 2001
Astrophys. J. Suppl. {\bf 135} 155.


\bibitem{Punsly} Punsly B. 1997, Astron. J. {\bf 114}, 544.


\bibitem{Reimer} Reimer O., Brazier K. T. S.,
 Carramiñana, A., Kanbach G., Nolan P. L., \&
 Thompson D. J., 2001
Mon.\ Not.\ Roy.\ Astron.\ Soc.\  {\bf 324}, 772.


\bibitem{Sigl:2000sn}
Sigl G., Torres D. F., Anchordoqui L. A., \& Romero G. E., 2001
Phys.\ Rev.\ D {\bf 63}, 081302.


\bibitem{Stanev:1995my}
Stanev T., Biermann P. L., Lloyd-Evans J., Rachen J. P. \& Watson A., 1995
Phys.\ Rev.\ Lett.\  {\bf 75}, 3056.


\bibitem{SE} Sowards-Emmerd D., Romani R.W., Michelson P.F. 2002, Astrophys. J. {\bf 509}, 109.

\bibitem{Tinyakov:2001nr}
Tinyakov P. G. \& Tkachev I. I., 2001
JETP Lett.\  {\bf 74}, 445
[Pisma Zh.\ Eksp.\ Teor.\ Fiz.\  {\bf 74}, 499 (2001)]



\bibitem{Tinyakov:2001ir}
Tinyakov P. G. \& Tkachev I. I., 2002
Astropart.\ Phys.\  {\bf 18}, 165.


\bibitem{Tinyakov:2003bi}
Tinyakov P. \& Tkachev I., 2003
arXiv:astro-ph/0301336.



\bibitem{Torres:2002bb}
Torres D. F., Boldt E., Hamilton T. \& Loewenstein M., 2002
Phys.\ Rev.\ D {\bf 66}, 023001.

\bibitem{Torres:2002af}
Torres D. F., Romero G. E., Dame T. M., Combi J. A. \& Butt Y. M,
2003a
Phys. Rept. {\bf 382}, 303.

\bibitem{Towers} Torres D.F. 2003b, to appear in "Cosmic Gamma-ray Sources", Kluwer University Press,
edited by K.S. Cheng and G.E. Romero.

\bibitem{Uchihori:1999gu}
Uchihori Y., Nagano M., Takeda M., Teshima M., Lloyd-Evans J. \& Watson A. A., 2000
Astropart.\ Phys.\  {\bf 13} 151.


\bibitem{Veron} Veron-Cetty M. P. \& Veron P., 2000
{\it A catalogue of Quasars and Active Galactic Nuclei, 9th Edition,}
ESO Scientific Report.

\bibitem{Veron:AA} Veron-Cetty M. P. \& Veron P., 2001
Astron. Astrophys. {\bf 374}, 92.

\bibitem{Virmani:xk}
Virmani A., Bhattacharya S., Jain P., Razzaque S., Ralston J. P. \& McKay D. W., 2002
Astropart.\ Phys.\  {\bf 17}, 489.

\bibitem{Weiler:1997sh}
Weiler T.~J., 1999
Astropart.\ Phys.\  {\bf 11}, 303.

\bibitem{Zatsepin:1966jv}
Zatsepin G. T. and Kuzmin V. A., 1996
JETP Lett.\  {\bf 4}, 78
[Pisma Zh.\ Eksp.\ Teor.\ Fiz.\  {\bf 4}, 114 (1966)].

\end{thebibliography}
\end{document}